\documentclass[a4paper,11pt]{article}

\usepackage[dvips]{graphicx}
\usepackage{amsmath,amssymb}
\usepackage{color}

\usepackage[normalem]{ulem}

\newcommand{\bX}{\boldsymbol{X}}
\newcommand{\bx}{\boldsymbol{x}}

\newcommand{\by}{\boldsymbol{y}}

\newcommand{\bM}{\boldsymbol{M}}

\newcommand{\be} {\begin{equation}}
\newcommand{\ee}{\end{equation}}

\begin{document}

\author{Adam B.~Barrett*$^1$, Pedro A.~M.~Mediano$^2$\\\\
$^{1}$\textit{ Sackler Centre for Consciousness Science} and \\
\textit{Department of Informatics}, University of Sussex, Brighton, UK\\
$^2$\textit{Department of Computing}, Imperial College, London, UK\\
\\
*adam.barrett@sussex.ac.uk (correspondence)
}

\title{The $\Phi$ measure of integrated information is not well-defined for general physical systems}

\date{}

\maketitle

\begin{abstract}
\noindent According to the Integrated Information Theory of Consciousness, consciousness is a fundamental observer-independent property of physical systems, and the measure $\Phi$ of integrated information is identical to the quantity or level of consciousness. For this to be plausible, there should be no alternative formulae for $\Phi$ consistent with the axioms of IIT, and there should not be cases of $\Phi$ being ill-defined. This article presents three ways in which $\Phi$, in its current formulation, fails to meet these standards, and discusses how this problem might be addressed.
\end{abstract}

\section{Introduction}

A key component of integrated information theory (IIT) is the mathematical formalism for supposedly describing quantitatively the extent and nature of the consciousness (subjective experience) generated by any physical system (Oizumi et al, 2014). The theory claims that at every moment that the physical state is updated, there is potential for a conscious experience to be generated. The ``intrinsic informational structure" of the mechanisms behind a state transition governs the quality of the experience, whilst the overall quantity of consciousness generated is identical to the system's value of the measure $\Phi$ (Tononi et al, 2016).  The quantity $\Phi$ essentially captures the extent to which the whole system is generating intrinsic information over and above its parts. By \textit{intrinsic} information it is meant that which is independent of the frame of reference imposed by outside observers of the system. The axioms and postulates of IIT state that consciousness is a fundamental, observer-independent property of physical systems, analogous to mass, charge or energy (Tononi and Koch, 2015), and hence imply that $\Phi$ is a fundamental physical quantity.

Much of the critique of the $\Phi$ measure has been based on the impracticality of its application to empirical neural data, and thus its inability to make testable predictions for IIT (e.g. Bor, 2012). Notably, the computation time required to compute $\Phi$ grows faster than exponentially with the number of system components;\footnote{More precisely, the computation time required to compute the effective $\Phi$ for a particular system graining grows faster than exponentially with the number of components. To obtain the maximum $\Phi$ over all grainings is intractable in the absence of a short-cut. See Section 2.1.} and it has only ever been computed on a specific kind of toy model system with just a handful of components (Mayner et al, 2017). Here we set testability issues aside, and address the deeper question of whether it it is theoretically possible for $\Phi$ to be a fundamental physical quantity. For it to be so, it must be well-defined, and there should be no alternative formulae for $\Phi$ consistent with the axioms and postulates of the theory. Here, we list three ways in which it is not well-defined, and hence conclude that further development of the theory and operationalisation of $\Phi$ is required.

%We do not here go into all the details of how to compute $\Phi$, but just describe the essential steps that are needed for the argument below. 

\section{Key quantities for the construction of $\Phi$}
This section provides some description of the construction of $\Phi$ in words, and writes down the key mathematical quantities (probability distributions) from which it is constructed.  A detailed description is not provided; for that the reader is referred to Oizumi et al (2014). $\Phi$ has been developed and illustrated via the use of examples of toy model systems consisting of indivisible and discrete binary components (logic gates). These systems evolve in discrete time; at each discrete time-step each component has its state updated according to the specified interactions (mechanisms) present. The dynamics are memoryless (Markovian): the probability distribution for the state at the next time-step only depends on the present state, and not on the past history. Most real complex systems are not easily modelled in this way, and this is a source of the theoretical problems presented below. Defining $\Phi$ relies on quantifications of (i) information and (ii) integration, in the system. Each of these components is rehearsed here in turn. 

Information is specified as that which the current state of the system contains about a hypothetical past state in which all configurations of the system were \textit{a priori} equally likely.\footnote{Only integrated information of `cause' is considered here. For integrated information of `effect' one swaps $t=0$ and $t=1$. The final $\Phi$ is the minimum of that computed for causes and that computed for effects.} Informally, the concept is that the more past states that are ruled out (or made improbable) by the current state, the greater the information generated. Formally, the key quantity is the joint probability distribution $P_{\mathrm{ce}}(\bX_0,\bX_1)$ for the states $\bX_0$ and $\bX_1$ of the system at discrete time $t=0$ and $t=1$, given that the system was perturbed at $t=0$ into all possible states with equal probability (Krohn and Ostwald, 2017). The acronym `ce' stands for cause-effect. This quantity decomposes as
\be
P_{\mathrm{ce}}(\bX_0,\bX_1)=P(\bX_1 | \bX_0) P_{\mathrm{u}}(\bX_0)\,, \label{eq: Pce}
\ee
where $P_{\mathrm{u}}(\bX_0)$ is the uniform (or \textit{maximum entropy}) distribution and $P(\bX_1 | \bX_0)$ is given by the system's dynamics. From this, the conditional distribution $P_{\mathrm{c}}(\bX_0 | \bX_1)$ is extracted:
\be
P_{\mathrm{c}}(\bX_0=\bx_0 | \bX_1=\bx_1)=:\frac{P_{\mathrm{ce}}(\bx_0,\bx_1)}{\sum_{\bx^*} P_{\mathrm{ce}}(\bx^* ,\bx_1)}\,.
\ee
As discussed below: (i) $P(\bX_1 | \bX_0)$, and hence $P_{\mathrm{ce}}(\bX_0,\bX_1)$, is only well-defined for Markovian systems; (ii)   $P_{\mathrm{u}}(\bX_0)$ is only defined if the set of states is finite (or else compact, i.e.~closed and bounded). 

Integration is operationalised by comparing probability distributions associated with the whole system to analogous probability distributions associated with a partition of the system. For the comparison, the probability distributions associated with distinct parts within a partition are taken to be independent. Formally, one computes the distance in probability distribution space between the probability distribution for the whole and the product of the probability distributions for the parts. To define the parts, one must specify a \textit{partition} $\mathcal{P} = \{M^1, M^2, \dots, M^r\}$ that divides the elements of $X$ into $r$ non-overlapping, non-trivial sub-systems, such that $X=M^1 \cup M^2 \cup \ldots \cup M^r$. With these key elements defined, integration is quantified by considering the distance between
\be
P_{\mathrm{c}}(\bX_0 | \bX_1=\bx) \hspace{0.3cm} \mathrm{and} \hspace{0.3cm} \prod_k P_{\mathrm{c}}(\bM^k_0 | \bM^k_1 =\boldsymbol{m}^k)
\ee
where the $\boldsymbol{m}^k$ are the sub-system states corresponding to whole system state $\bx$ under the partitioning. The greater the distance between these, in probability distribution space, the greater the amount of integrated information (with respect to the given partition). The metric on probability distribution space is taken to be the ``earth mover's'' (or Wasserstein) distance (Oizumi et al, 2014). Then $\Phi$ is the minimum of this distance taken across all possible partitions -- in what is commonly known as taking the ``cruelest cut'' of the system; see Oizumi et al (2014) and Krohn and Ostwald (2017) for details.

\subsection{Maximisation over possible grainings} \label{sec:grainings}

Importantly, to compute $\Phi$, a graining of the system is needed, in space, time and the set of possible states of the components. Since the measure is supposed to be independent of the point of view of the observer, the choice of grainings must be observer-independent. It is prescribed that the grainings to be used are those that lead to the maximum possible value of $\Phi$. Thus, the $\Phi$ of a graining is the minimum across partitions of that graining, and the final $\Phi$ of the system is the maximum over all possible grainings.

This maximisation over grainings is currently infeasible to carry out in practice for any real physical system, since no compelling short-cuts or approximations yet exist for searching through the infinity of possibilities (Barrett, 2016). Hence computation of $\Phi$ is currently intractable for any real physical system. Nevertheless, from the intrinsic (or ontological) perspective, a physical system may instantiate its own maximisation despite that maximisation being infeasible to compute by any external observer.  The problems highlighted in Section 3 are distinct from and go beyond this practical computability challenge. We emphasise this issue of graining here since the requirement to maximise over all grainings increases the extent to which $\Phi$ is not well-defined, according to Problems 2 and 3 below.

\section{Three ways in which $\Phi$ is not well-defined} 

This section lists three ways in which $\Phi$, as currently formulated, is not well-defined. 

%
%These models differ greatly in their physics from real systems. Physical information processing systems such as real or artificial neural networks are made of physical matter and the apparent components (neurons, transistors etc.) are divisible into conglomerations of molecules, atoms, quarks and leptons. 

%Computing $\Phi$ involves considering discretisations of the system in question, with grainings applied in space, time and to the set of possible states of the components. Current versions of IIT propose that the grainings to be used are those that lead to the maximum possible value of $\Phi$. (This leads to it being impractical to compute; however this problem itself doesn't rule out the physicality of $\Phi$, as there are many things we can't compute in this universe.) Given a graining of a system, $\Phi$ is built from the 

\subsection{Problem 1: There is no canonical metric on the space of states, nor a canonical metric on probability distribution space}
In order to compute the earth mover's distance, a metric is required on the space of states, i.e.~one requires there to be a well-defined distance between any two states. In IIT-3.0 the Hamming distance is proposed as the distance when the state of each component of the system is binary. However, for general non-binary states there are a range of possible metrics and no canonical `intrinsic' choice. For two states $\bx$ and $\by$, example valid expressions for the distance between them are the `L1 norm' $\sum_i | x_i-y_i |$, the `L2 norm' $\sqrt{\sum_i ( x_i-y_i )^2}$ and the `$L\infty$ norm' $\max_i | x_i -y_i |$. The earth mover's distance is further not the only distance measure for probability distributions. Tegmark (2016) lists several alternatives, and there is no canonical choice.  As demonstrated in simulation in Mediano et al (2018), different choices in the construction of a variant $\Phi$ measure can lead to profound differences in the behaviour, even on small systems. Therefore, in the absence of a well-argued principle by which to uniquely fix the metrics on the space of states and on probability distribution space, $\Phi$ is not well-defined. 

\subsection{Problem 2: The requirement of a discrete set of states}
The maximum entropy distribution on the set of states of the system is not well-defined if the set of states is infinite (except for the case of the set of states being compact, i.e.~closed and bounded [Barrett and Seth, 2011]). Thus, for example, the neuron membrane potential can not be taken as the state variable, since there is no way to define precise absolute limits on it. More generally, any system with Gaussian or exponentially distributed state variables does not have a well-defined $\Phi$ for this reason. A possible fix might be to only consider grainings into discrete sets of states. However, one would need a canonical method for labelling a discrete set of states obtained from the continuous variable, and to have solved Problem 1 above for there to then be a canonical metric on any discretisation. Then, one would need to show that there always exists an upper bound to the effective $\Phi$ over all discrete finite grainings.

\subsection{Problem 3: The requirement of Markovian dynamics}
Additionally, $\Phi$ is not well-defined for a system with non-Markovian dynamics, i.e. one for which the dynamics are not memoryless (Barrett and Seth, 2011). The probability distributions $P(\bX_1 | \bX_0)$ and $P_{\mathrm{ce}}(\bX_0,\bX_1)$ in the formula are not well-defined unless the probability distribution for future states depends only on the current state, and not on the system's past history. This is because for a non-Markovian system the distribution on the past history given $\bX_0$ is not specified. We highlight that this is an important problem, and not just merely a theoretical construct -- brain dynamics are non-Markovian at many levels, ranging from the EEG level (see for example von Wegner et al, 2017), to the level of ionic current fluctuations in membrane channels (see for example Fuli\'{n}ski et al, 1998). More generally, a system may be Markovian with respect to some grainings, but not for all grainings. Given that $\Phi$ is supposed to be specified by the maximisation over all possible grainings, (see Section \ref{sec:grainings}), it only takes one graining of a given system to have non-Markovian dynamics for $\Phi$ to be ill-defined for that system.

For non-Markovian systems, one might perhaps attempt to define $\Phi$ as the limit as $k\to\infty$ of the analogous quantity with $\bX_0$ replaced everywhere with $(\bX_0, \bX_{-1}, \bX_{-2}, \ldots, \bX_{-k})$, i.e.~try setting all past states in an indefinitely long past history to be independent and maximum entropy under the perturbation, and see if there is convergence as the length of past history considered tends to infinity. Such an approach would not however solve the issue for non-ergodic systems: for a non-ergodic system, by definition, there is no convergence of $P(\bX_1|\bX_{0},\bX_{-1},\bX_{-2},\ldots,\bX_{-k})$.

As an example, we consider a non-ergodic system, which has non-Markovian grainings. This system has a variable $S$ (potentially in addition to other variables, which we need not consider to make the point), which follows a random walk:
\be
S_{t+1}=S_t+B_t\,,
\ee
where the $B_t$ are independent identically distributed binary random variables with equal probability of taking the values -1 and 1. Consider the binarisation of this variable $S$ such that the binary state $X$ is given by $X=1$ if $S$ exceeds some threshold $\theta$, and $X=0$ otherwise. To compute $\Phi$ for this graining we would need the quantity 
\be
P(X_1=1 | X_0=1)=P(S_1 > \theta | S_0 > \theta)=\sum_{s=\theta+1}^{\infty} P(S_1 > \theta | S_0 =s)P(S_0=s)\,.
\ee
But this is not well-defined since there is no well-defined probability density function for $S_0$: one cannot impose a maximum entropy distribution because the range of values $S_0$ can take is not a compact set (see Problem 2 above), because as the length of history considered tends to infinity, the set of possible values of $S_0$ goes to infinity also.

%It is unclear whether this would work for all possible grainings of all possible systems. A single example physical system for which convergence doesn't happen would rule out this approach. 

\section{Discussion}
For IIT to mature as a theory, the three problems above will need to be addressed. This article concludes with some discussion on this. 

Problems 2 and 3 arise from needing to quantify the information that the current state holds about some prior state. The maximum entropy distribution is the only possible prior one can impose on the past state, as any other choice will depend on some arbitrary information held by the observer. An empirical distribution can not be used, because not all systems are stationary: the statistics of the system could change the moment any recording is terminated, so one would never know if one has recorded everything that the system could have done. A reformulation of $\Phi$ in terms solely of the geometrical and topological structure of the instantaneous state of the system, without reference to past and future states, might be the only way of solving Problems 2 and 3 (Barrett, 2014). This would change the fundaments of the theory somewhat- it would be less about the mechanisms underlying the evolution of the system, and more about simply obtaining a mapping from a physical structure onto the structure of the phenomenal experience associated with the physical structure. Nevertheless, a complex set of mechanisms are needed to generate a system that can exist in multiple complex configurations, so mechanisms would still in some sense be fundamental to consciousness on such an updated theory. 

Problem 1 appears to be harder to solve. However one might reformulate the theory, any attempt to create a formula for consciousness as intrinsic information needs to define, spatially, where one system ends and another begins. Without a canonical metric on the space of system configurations, one would not be able to quantify differences between systems and sub-systems in a truly observer-independent fashion. It might be that the possible metrics are heavily constrained by the requirement that the effective $\Phi$ must always remain bounded under increasingly fine grainings (see Problem 2); such an investigation is beyond the scope of this paper, but could form the basis for future work.

%An approach based not on discretization, but rather on continuous fields evolving in continuous time would obviate the need to consider alternative grainings of states, or system components (Barrett, 2016). Such an approach might link up better with fundamental physics, given that fields (e.g.~the electromagnetic field) are considered fundamental in standard fundamental models in physics (Barrett, 2014). 

%A formulation capturing in an observer-independent way the geometrical and topological structure of the instantaneous (electromagnetic) field configuration, without reference to past and future states, might further obviate the need to consider a maximum entropy distribution, and hence solve problems 1 and 2.

Successful observer-independent theories for how macroscopic physics emerge from fundamental entities are typically cast in terms of continuous fields, e.g. Einstein's theory of mass and gravitation (general relativity), and Maxwell's theory of electromagnetism (Barrett, 2016). Barrett (2014) proposes that an approach to IIT, and the emergence of consciousness, based on fields might offer advantages over the existing discretization-based approach. It is a debatable supposition that the state of consciousness of a physical system is determined by its structure at a variable spatio-temporal scale and state graining, given by that which happens to maximise $\Phi$ for the given system at the given moment (Bayne, 2018). If a formula for the integrated information intrinsic to a field configuration could be obtained, there would be no need to consider alternative grainings of states, or system components. Because human consciousness arises from complex electrical activity in the brain, the hypothesis would be that its fundamental substrate is the integrated information intrinsic to specifically the electromagnetic field (as opposed to say, the gravitational or nuclear force field) it generates (Barrett, 2016); see Barrett (2014) for more on this idea. 

Continuing to attempt a formulation of intrinsic information via discrete graining, one might make use of quantities related to Kolmogorov-Sinai (KS) entropy (Sinai, 2009). KS entropy is well-defined for all ergodic systems as a supremum over all grainings. Furthermore, Thurner and Hanel (2012) recently proposed a formalism for defining generalised entropies for non-ergodic systems. Perhaps that could be used to generalise KS entropy to non-ergodic systems, and hence to obtain a universally well-defined intrinsic description of information dynamics.

\subsection{Final remarks}

We have shown that the supposedly fundamental $\Phi$ measure of integrated information, as described in IIT version 3.0 (Oizumi et el., 2014) is not well-defined for general physical systems. We have not addressed here the many variant $\Phi$ measures that have been developed for potential practical application to specific classes of systems, see Tegmark (2016) and Mediano et al. (2018) for reviews. These tend to quantify information with respect to the empirical distribution as opposed to the maximum entropy distribution, and can be applied to systems with continuous states (Problem 2 doesn't apply), and moreover to any stationary system (Markovian or not). Further, for a non-linear deterministic system, a distinct approach to operationalising integrated information in terms of topological dimensionality of attractor dynamics has been proposed (Tajima and Kanai, 2017). Any of these measures might be tested for \textit{correlation} with consciousness when computed across choice sets of brain variables (Barrett and Seth, 2011). However, the behaviour of these various measures is very diverse even on small simple networks, so one must remain cautious about considering them as generalisations or approximations of any eventual, `fundamental' $\Phi$ measure (Mediano et al., 2018).

The key idea of IIT, that consciousness is, in some sense, intrinsic information remains intriguing and influential (Tegmark, 2015). However, operationalising this idea and obtaining a candidate universal mathematical description of intrinsic information remains challenging. The current $\Phi$ measure is neither universally well-defined, nor fully independent of certain arbitrary choices input into its construction. It is in the best interest of IIT that we recognise and address these problems to move towards a truly plausible measure of phenomenal experience from physical structure.

%And further, this introduces the problem of granularity in time. 

%If we found a new formulation that was based only on the structure of the instantaneous state of the system, then we might be able to solve these problems. In CITE Barrett it was proposed that an approach to IIT based on continuous fields might offer advantages over the existing discretization-based approach. Further, fields are actually fundamental entities in modern physics. (This doesn't mean we need to quantum field theory. To describe structure at the spatiotemporal scale of neurodynamics, Maxwell's classical field equations for electromagnetism are a good fit.) If a formula were obtained for the integrated information intrinsic to a field configuration, there would be no need to consider alternative grainings of states, or system components. 
%
%A reformulation of IIT based only on the instantaneous state of the system could potentially obviate the need to utilise a maximum entropy distribution. 
%
%The hardest problem to solve seems to be that of finding a metric for comparing a whole system to a subdivided system. 
%
%Empirical Phi measures for real world work. But empirical Phi measures no good for being a fundamental measure. 
%\\
%
%Categorisation of problems into epistemological and ontological?

%Thus, even theoretically, it cannot in its current form be a fundamental physical quantity. If we wish to maintain the idea that consciousness can essentially be identified with intrinsic information, then a radically different approach will be needed.
 
 \section*{Acknowledgements}
 ABB is funded by EPSRC grant EP/L005131/1.

\section*{References}

Barrett, A.B. (2014) An integration of integrated information theory with fundamental physics, \textit{Front. Psychol.}, 5, 63.\\

\noindent Barrett, A.B. (2016) A comment on Tononi \& Koch (2015) `Consciousness: here, there and everywhere?', \textit{Phil. Trans. R. Soc. B}, 20140198.\\

\noindent Barrett, A.B., \& Seth, A.K. (2011). Practical measures of integrated information for time-series data. \textit{PLoS Comput. Biol.}, 7(1): e1001052.\\

\noindent Bayne, T. (2018). On the axiomatic foundations of the integrated information theory of consciousness, \textit{Neuroscience of Consciousness}, 2018(1), niy007.\\

\noindent Bor, D. (2012) \textit{The Ravenous Brain: How the New Science of Consciousness Explains Our Insatiable Search for Meaning}, New York, NY: Basic Books.\\ 

\noindent Fuli\'{n}ski, A., Grzywna, Z., Mellor, I, Siwy, Z.,  \& Usherwood, P.N.R. (1998) Non-Markovian character of ionic current fluctuations in membrane channels \textit{Phys. Rev. E} 58, 919.\\

\noindent Krohn, S. \& Ostwald, D. (2017) Computing Integrated Information, \textit{Neuroscience of Consciousness},  2017(1), nix017.\\

\noindent Mayner, W.G.P., Marshall, W., Albantakis, L., Findlay, G., Marchman, R. \& Tononi, G. (2017) PyPhi: A toolbox for integrated information theory,  \textit{arXiv}, 1712.09644.\\

\noindent Mediano, P.A.M., Seth, A.K., \& Barrett, A.B. (2018). Measuring integrated information: Comparison of candidate measures in theory and simulation. \textit{arXiv}, 1806.09373.\\

\noindent Oizumi, M., Albantakis, L. \& Tononi, G. (2014) From the phenomenology to the mechanisms of consciousness: Integrated Information Theory 3.0, 
\textit{PLoS Computational Biology}, 10 (5), e1003588.\\

\noindent Sinai, Y. (2009) Kolmogorov-Sinai entropy,
\textit{Scholarpedia}, 4(3):2034.\\

\noindent Tajima, S., \& Kanai, R. (2017) Integrated information and dimensionality in continuous attractor dynamics, \textit{Neuroscience of Consciousness}, 2017(1), nix011.\\

\noindent Tegmark, M. (2015) Consciousness as a state of matter, \textit{Chaos, Solitons and Fractals}, 76, 238-270.\\

\noindent Tegmark, M. (2016) Improved Measures of Integrated Information. \textit{PLoS Comput Biol}, 12(11): e1005123.\\

\noindent Thurner, S. \& Hanel, R. (2012) The entropy of non-ergodic complex systems - A derivation from first principles. \textit{International Journal of Modern Physics: Conference Series}, 16, 105-115.\\

\noindent Tononi, G. \& Koch, C. (2015) Consciousness: here, there and everywhere? \textit{Phil Trans. R. Soc. B} 370, 20140167.\\

\noindent Tononi, G., Boly, M., Massimini, M., \& Koch, C. (2016) Integrated information theory: from consciousness to its physical substrate \textit{Nature Reviews Neuroscience} 17, 450-461.\\

\noindent von Wegner F., Tagliazucchi, E., \& Laufs, H. (2017) Information-theoretical analysis of resting state EEG microstate sequences - non-Markovianity, non-stationarity and periodicities. Neuroimage. 158, 99-111.

\end{document}